\documentclass{article}
\usepackage{amsmath}
\usepackage{authblk}
\usepackage{float}
\usepackage[a4paper, total={6in, 8in}]{geometry}
\usepackage{graphicx} 
\usepackage{hyperref}
\usepackage{natbib}

\title{Lowering In-Memory Footprint of Antenna Beams via Polynomial Approximation}

\author[1]{Ali Taqi}
\author[1]{Karel Ad\'amek}
\author[2]{Quentin Gueuning}
\author[3]{Maciej Serylak}
\author[3]{Robert Laing}
\author[1]{Wesley Armour\footnote{Email address: wes.armour@oerc.ox.ac.uk}}

\affil[1]{\footnotesize Oxford e-Research Centre (OeRC), Dept. of Engineering Science, Univ. of Oxford, OX1 3QG, UK}
\affil[2]{\footnotesize Battcock Centre for Experimental Astrophysics, Cavendish Laboratory, Univ. of Cambridge, CB3 0HE, UK}
\affil[3]{\footnotesize SKA Observatory, Jodrell Bank, SK11 9FT, UK}

\date{December 2024\footnote{Pre-print draft version}}

\begin{document}

\maketitle

\begin{abstract}
    With the emergence of new radio telescopes promising larger fields of view at lower observation frequencies (e.g., SKA), addressing direction-dependent effects (DDE) (e.g., direction-specific beam responses), polarisation leakage, and pointing errors has become all the more important. Be it through A-projection or otherwise, addressing said effects often requires reliable representations of antenna/station beams; yet, these require significant amounts of computational memory as they are baseline-, frequency-, time-, and polarisation-dependent. A novel prototype is reported here to approximate antenna beams suitable for SKA-MID using Zernike polynomials. It is shown that beam kernels can be well approximated, paving the way for future optimisations towards facilitating more efficient beam-dependent solutions and approaches to tackling the aforementioned challenges, all of which are essential for large-scale radio telescopes.
\end{abstract}

\section{Introduction}
High-resolution, wide-field radio telescopes promise major scientific breakthroughs; however, they also demand tackling various challenges such as direction-dependent effects (DDE), polarisation leakage, and pointing errors. Addressing such phenomena typically demands effective characterisation of antenna beams, which are not only polarisation-dependent, but also depend on baseline, frequency, and time. Hence, novel beam and pointing correction approaches (e.g., A-projection \citep{2013ApJ} and pointing selfcal \citep{2017AJ}, which incorporate beam kernels into radio imaging pipelines on-the-fly, may reserve significant amounts of computational memory. Therefore, reducing such computational demand via analytical beam approximation would address a key bottleneck in data processing, enabling efficient implementation of beam-dependent solutions for novel radio telescopes.

This paper aims (i) to evaluate via a prototype whether Zernike polynomials can approximate antenna beams; (ii) to understand if and where said approximations can be computationally intensive.

\section{Zernike polynomials}
These are an array of orthogonal polynomials on a unit disk, initially developed to describe diffraction patterns in phase contrast imaging. The formal definition under the Noll indexing scheme is given in Equations 1 \& 2.

\begin{equation}
    Z_{j}(r, \theta) = Z_{mn}(r, \theta) = \begin{cases} 
      \sqrt{2(n+1)}R_{mn}(r)\cos{m\theta} & m \neq 0; j = even \\
      \sqrt{2(n+1)}R_{mn}(r)\sin{m\theta} & m \neq 0; j = odd \\
      \sqrt{n+1}R_{mn}(r) & m = 0 
   \end{cases}
\end{equation}

\begin{equation}
    R_{mn}(r) = \sum_{s=0}^{(n-m)/2} \frac{(-1)^{s}(n-s)!}{s!(\frac{n + m}{2} - s)!(\frac{n - m}{2} - s)!} r^{n-2s}
\end{equation}

\begin{figure}[H]
    \centering
    \includegraphics[width=0.5\textwidth]{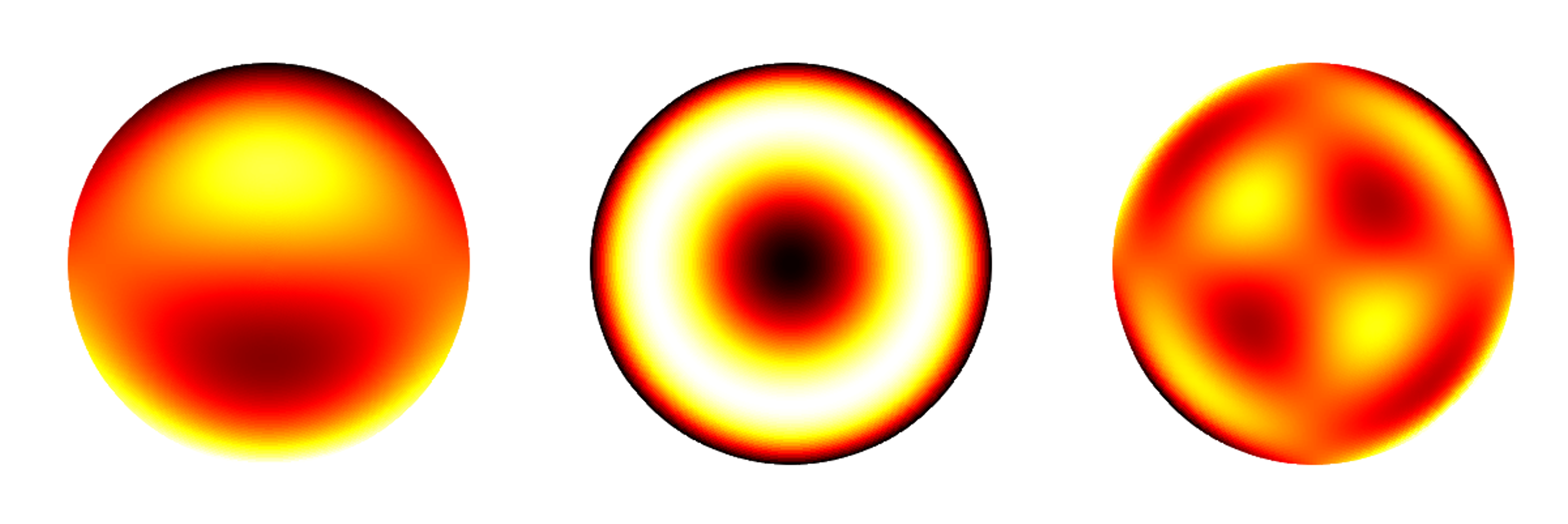}
    \caption{
        \centering
        \small{Examples of Zernike aberrations: left to right - $j=7$; $j=11$; $j=23$}
    }
    \label{fig1}
\end{figure}

$Z(r, \theta)$ is the Zernike value at a given radius $r$ and angle $\theta$. $j$ and $(m, n)$ represent aberration indices and are interconvertible via Equation 3. Figure \ref{fig1} visualises some Zernike aberrations in 2D. Note that the current scheme only supports real numbers.

\begin{equation}
    j = \begin{cases}
        \frac{n(n + 1)}{2} + 1 & m = 0 \\
        \left[ \frac{n(n + 1)}{2} + m, \frac{n(n + 1)}{2} + m + 1 \right] & m \neq 0
    \end{cases}
\end{equation}

\section{Prototype with MeerKAT beams}

\subsection{Methodology}
Experimental MeerKAT beams were obtained from \citet{2023AJ}, and for this prototype, two beams representing the \textit{xx} and \textit{xy} polarisations (magnitude only) were utilised as demonstrated in Figure \ref{fig2}. Approximating these beams entailed fitting several Zernike aberrations using a non-linear least-squares method, generating analytical approximations comprising a collection of weighed and stacked Zernike aberrations in each case.

\subsection{\textit{xx} polarisation}
\begin{figure}[H]
    \centering
    \includegraphics[width=0.38\textwidth]{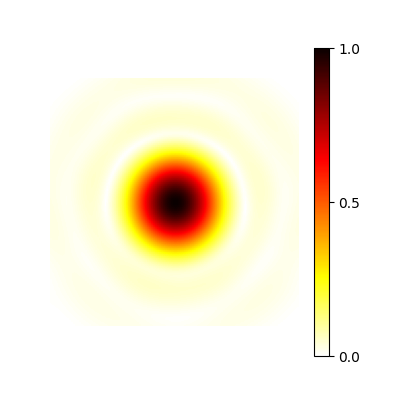}
    \includegraphics[width=0.38\textwidth]{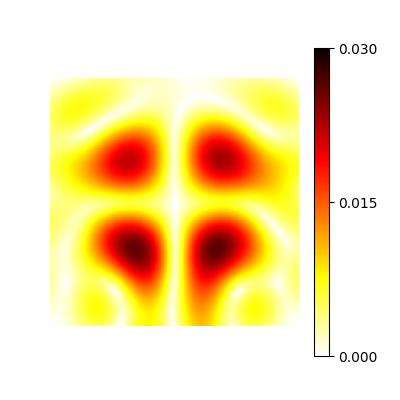}
    \caption{
        \centering
        \small{MeerKAT beam magnitudes; L-band @ 857 MHz; antenna 0; polarisations: \textit{left to right} - \textit{xx}; \textit{xy} \citep{2023AJ}; license: \href{https://creativecommons.org/licenses/by-nc/4.0/}{CC BY-NC 4.0} - \url{https://creativecommons.org/licenses/by-nc/4.0/}}
    }
    \label{fig2}
\end{figure}

\begin{figure}[H]
    \centering
    \includegraphics[width=0.8\textwidth]{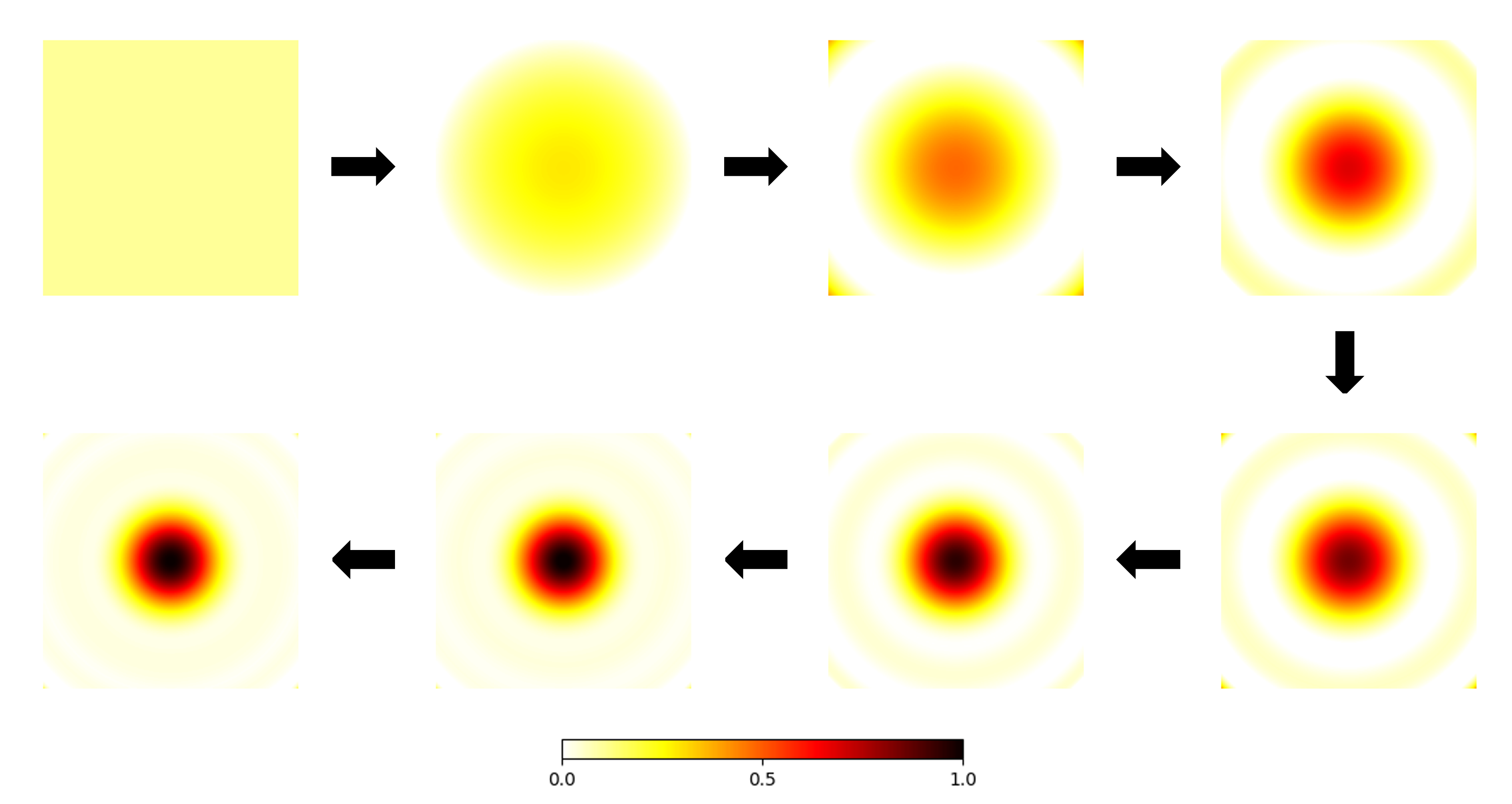}
    \caption{
        \centering
        \small{Beam approximations for the \textit{xx} polarisation case. Incorporated Zernike aberrations (incrementally): $j=1$; $j=4$; $j=11$; $j=22$; $j=37$; $j=56$; $j=79$; $j=106; 137; 172; 211; 254; 301; 352; 407; 466$}
    }
    \label{fig3}
\end{figure}

Given the regular, Gaussian-like shape of the beam in this case (Figure \ref{fig2} - \textit{left}), only central aberrations lying on the altitude of the Zernike triangle were used. These were added incrementally as shown in Figure \ref{fig3}.

Convergence is evident as the maximum error between the estimated and experimental beams drops dramatically from $0.9 units$ for $j=1$ to $0.1 units$ with $j \leq 79$. Such convergence is achievable with just $7$ weighed Zernike aberrations, signifying reasonably low memory demands due to the relatively small number of computations required for Equations 1 \& 2.

\subsection{\textit{xy} polarisation}
\begin{figure}[H]
    \centering
    \includegraphics[width=0.8\textwidth]{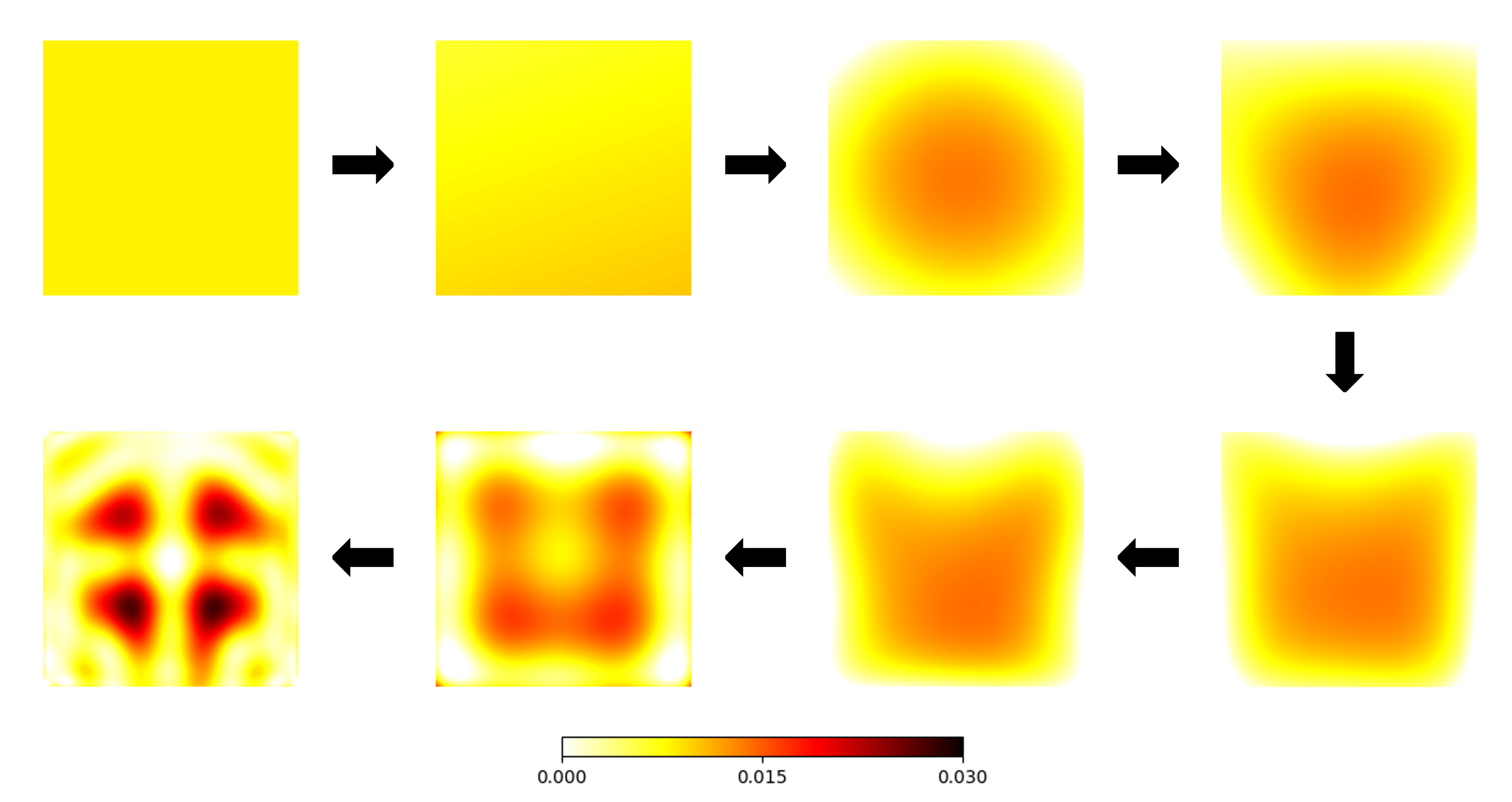}
    \caption{
        \centering
        \small{Beam approximations for the \textit{xy} polarisation case. Incorporated Zernike aberration rows (incrementally): $n=0$; $n=1$; $n=2$; $n=3$; $n=4$; $n=5$; $n=6$; $n=7 - 15$}
    }
    \label{fig4}
\end{figure}

Given the more complex beam in this case (Figure \ref{fig2} - \textit{right}), all Zernike aberrations up to some limit were considered for the estimation. Figure \ref{fig4} shows the results where the Zernike aberrations used were clustered into $n$ rows matching the Noll indexing scheme, and each row was added incrementally.

While convergence is also evident here, it is significantly more demanding computationally as it requires an order of magnitude more aberrations. Future work should optimise this process through: (i) adopting a more informed selection mechanism for aberrations based on their underpinning scientific suitability to the use case; (ii) implementing a different scheme for Zernike polynomials, preferably one which accounts for complex numbers thus enabling more comprehensive analysis but also offering potential mathematical simplifications and further memory savings.

\section{Outcomes and forward direction}
Zernike polynomials can estimate antenna beams with high precision, promising a memory-efficient solution for addressing beam-dependent DDE, polarisation leakage, and pointing errors. Nonetheless, computational demand for these estimations should be further reduced through adopting more scientifically-informed aberration selection methods; and implementing a more efficient Zernike scheme preferably one allowing complex quantification. Future work should also focus on sensitivity analyses spanning various bands, frequencies, and polarisations, aiming to understand/extrapolate the fundamental effects of these parameters on the beam shape.

\small
\bibliographystyle{chicago}
\bibliography{references.bib}

\section*{Acknowledgments}
This work was supported by the Science and Technology Facilities Council of the UK [grant number ST/W001969/1].

\end{document}